\documentclass[11pt,a4paper]{article}
\usepackage{jheppub}
\title{Parisi-Sourlas-like dimensional reduction of quantum gravity in the
presence of observers}

\author[a]{Dmitriy Podolskiy,}

\author[b]{Andrei O. Barvinsky}

\author[c]{and Robert Lanza}

\affiliation[a]{Harvard University,\\
77 Avenue Louis Pasteur, Boston, MA 02115, USA}

\affiliation[b]{Lebedev Physics Institute, Theory Department\\
Leninsky Prospect 53, Moscow 117924, Russia}

\affiliation[c]{Wake Forest University,\\
1834 Wake Forest Rd., Winston-Salem, NC 27106, USA}

\emailAdd{Dmitriy\_Podolskiy@hms.harvard.edu}

\abstract{One of the sources of incompatibility between general relativity
and quantum mechanics is perturbative non-renormalizability of quantum
gravity in $3+1$ spacetime dimensions. Here, we show that in the
presence of disorder induced by random networks of observers measuring
covariant quantities (such as scalar curvature) $(3+1)$-dimensional
quantum gravity exhibits an effective dimensional reduction at large spatio-temporal scales, which is 
analogous to the Parisi-Sourlas phenomenon observed for quantum field theories
in random external fields. After averaging over associated disorder and focusing on the infrared dynamics of the theory we find that
the upper critical dimension of quantum gravity is lifted
from $D_{{\rm cr}}=1+1$ to $D_{{\rm cr}}=3+1$ dimensions.}


\notoc

\begin{document}
\maketitle
\flushbottom
\section{\label{sec:Intro}Introduction}
Difficulty of establishing connection between
general relativity and quantum mechanics has puzzled several generations
of theoretical physicists starting with Albert Einstein \cite{Einstein1935}.
At the heart of the problem is perturbative non-renormalizability of naively quantized
general relativity \cite{tHooft1974,Deser1974}: the theory becomes extremely sensitive to the choice of renormalization
scheme used essentially meaning that the perturbative control over the behavior of the theory is lost.

This problem resurfaces on multiple levels and within any physical
problem involving counting or accounting for quantum gravitational
degrees of freedom. For example, in perturbative calculation of gravitational
entropy associated with black hole horizon the numerical factor in
front of the horizon area acquires infinite perturbative corrections
(see \cite{Solodukhin2011} for the review) again strongly dependent
on the choice of regularization scheme thus entangling the information
loss paradox in quantum gravity \cite{Hawking1976} with the problem
of its non-renormalizability. In quantum cosmology, where vacuum energy
density essentially determines the expansion rate of the spacetime,
perturbative corrections to its value are strongly scale dependent
\cite{Weinberg1989}, making even the sign of vacuum energy density
hard to determine with certainty, and behavior of the theory in the
quantum cosmological setup is not controllable in both ultraviolet
and infrared limit.

Starting with the Weinberg's idea of asymptotic safety \cite{Weinberg1976},
it has been argued previously that canonical quantum gravity may be non-perturbatively
renormalizable, with a UV fixed point. Numerical simulations
of Regge-Wheeler simplicial quantum gravity (including the ones performed here, see \cite{Hamber1991} and references below),
simulations employing dynamical triangulations \cite{Ambjorn1992,Catterall1994,Bialas1996} as well as functional renormalization group 
analysis \cite{Reuter1996} indeed all point towards the validity of this conclusion.\footnote{Interestingly, simulations of both $4$-dimensional simplicial quantum gravity and dynamically triangulated $4$-dimensional quantum spacetime behave differently above and below the
UV fixed point with an AdS-like physics in the IR phase and a quasi-$2$-dimensional, branching polymer-like behavior in the UV phase.
While this observation is often used in the community as a reason to discard results of these numerical simulations --- IR behavior of spacetime
we live in is manifestly dS-like rather than asymptotically AdS one  --- we shall argue below that this difference in behavior above and below the fixed point is actually physical.}
However, given that gravity is the weakest force
(and, as it seems, would remain as such within any meaningful Grand Unification scheme \cite{ArkaniHamed2007}),
relying on the existence of a fixed point in a deeply UV regime feels unsatisfactory to us, as changing matter content
of the theory would change the location of the fixed point on its phase diagram possibly even removing it
altogether for specifically chosen matter content. Addressing the problem of non-renormalizability entirely within a perturbative domain, even 
given the extreme complexity of the problem, thus seems to us a more attractive possibility.

It has also became a common lore that a UV finite theory of gravity such
as string theory would automatically guarantee avoidance of the problem of non-renormalizability. 
Indeed, counting microstates associated with critical black
hole horizon in string theory gives a correct answer for numerical
prefactor in black hole entropy \cite{Strominger1996}. On the other
hand, it remains poorly understood (see \cite{Kaplan2020} as one of the most interesting latest works attempting to address this issue) if superstring theories provide
\emph{the} unique ultraviolet completion of naively quantized general
relativity (GR) or there might be other UV completions which lead
to the same controllable behavior in the continuum/infrared limit,
completions which we are currently not aware of. In the former case, there
should naturally exist a line of arguments which leads to emergence
of effective string theoretic representation of the ultraviolet physics
from an infrared effective GR setup, and it would be desirable to
demonstrate explicitly how a stringy behavior naturally emerges from
this setup in the UV limit. We believe that the present work identifies
a possible new research line along which such arguments can be obtained.

Namely, here we would like to argue that (a) including ``observers''
which continuously measure such covariant quantities as scalar curvature
(i.e., essentially probing the strength of gravitational interaction,
see below) and then averaging over disorder associated with a random
network of these observers and corresponding observation events leads
to an effectively de Sitter like behavior of the underlying theory
of quantum gravity, (b) deep infrared behavior of the resulting $3+1$-dimensional
theory is effectively reduced to the one of a $2$-dimensional theory,
and we identify a possible mapping between degrees of freedom
in the original, ($3+1)$-dimensional theory of quantum gravity (which
however includes disorder associated with observers, as was mentioned
above) and the ones in the effective $2$-dimensional quantum theory
obtained by averaging over disorder and taking the long wavelength
limit (such a mapping is introduced here at most in the first approximation as arguably the mapping dictionary we introduce below 
is far from being completely developed). The identified mapping is reminiscent of the celebrated Parisi-Sourlas
dimensional reduction known to take place in field theories with global
and gauge symmetries in the presence of random external fields \cite{Parisi1979}.
Finally, (c) we argue that the effective action of the emergent $2$-dimensional
theory coincides with the Liouville scalar theory, i.e., essentially,
the theory of two-dimensional quantum gravity \cite{Polyakov1981,Knizhnik1988}
possibly providing the missing link between naively quantized general
relativity and string theory and, importantly, a possible explanation
why observed dimensionality of spacetime which we live in is $D=3+1$.

We deem these observations interesting also because the described
setup, quantum gravity with disorder, represents a rare case in theoretical
physics when the presence of observers drastically changes behavior
of observable quantities themselves not only at microscopic scales
but also in the infrared limit, at very large spatio-temporal scales.
Namely, in the absence of observers the background of the $3+1$-dimensional theory remains unspecified. Once observers are introduced, 
coupled to the observable gravitational degrees of freedom and integrated out, the effective background of theory becomes de-Sitter like. Rather than being a
fundamental constant of the theory, the characteristic curvature of this de Sitter background spacetime (or effective cosmological constant) is determined by the intrinsic properties of observers such as the strength of their coupling to gravity and distribution of observation events across the fluctuating spacetime.
Physical observers represented by von Neumann detectors measuring
scalar curvature of spacetime (or other covariant quantities) play
a critically important role for our conclusions implying a necessity
of proper description of observer, observation event and interaction
between observers and the observed physical system for theoretical
controllability of the very physical setups being probed by observers.

The text of the manuscript is organized as follows. Section \ref{sec:Parisi-Sourlas-like-QG}
is devoted to a numerical study of simplicial Regge-Wheeler (Euclidean) quantum gravity
in the presence of random Gaussian field coupled to scalar curvature. We argue that the theory
exhibits an analogue of Parisi-Sourlas dimensional reduction after averaging over quenched disorder associated with events of gravitational field probing.
In Section \ref{sec:TheorArguments} we represent theoretical arguments
explaining results of this study and pointing towards their validity in a continuous
Lorenzian quantum theory. The Section \ref{sec:Conclusion}
is devoted to the outline of obtained results and a brief discussion of several analogies
of phenomena observed here and the ones realized in condensed matter physics.
Finally, appendices include details of numerical simulations of several quantum field theories 
which we used as a pilot study for the subsequent work on quantum gravity. They also contain a
more detailed theoretical derivation of the results of Section \ref{sec:TheorArguments}.

\section{Parisi-Sourlas-like dimensional reduction in Regge-Wheeler simplicial quantum gravity
\label{sec:Parisi-Sourlas-like-QG} }

Following the approach by Regge and Wheeler \cite{Regge2007,Wheeler1964,Hamber1984,Hamber1985,Hamber2000}, we consider
a pure $4$-dimensional Euclidean quantum gravity with a cosmological constant.\footnote{While Regge-Wheeler simplicial
gravity \cite{Hamber1985,Hamber2000} might very well be a very distant cousin of
the naively quantized general relativity, it is not yet entirely clear if (a)
the theory preserves local gauge invariance in the number of dimensions $D>2$ \cite{Hamber1997}, (b) Euclidean
setup critical for the theory is sufficient to capture essentially Lorentzian behavior of true Einstein gravity including,
in particular, its gravitational instability, and (c) the theory
actually contains a massless spin-2 particle in the spectrum of its low energy perturbations \cite{Hamber2004}. However, 
at the moment it remains the best setup which we can use attempting numerical studies of quantum general relativity.}

We are interested to determine possible changes in behavior of observables of the theory in the presence of
an extra ingredient: von Neumann observers randomly distributed across the fluctuating spacetime and measuring
the strength of gravitational self-interaction. Observational events associated with their
activity can be modeled by the term
\begin{equation}
\sqrt{g}\phi{}(x) R(x)=2\sum_{h\supset{x}}\phi_{x}\delta_{h}A_{h}
\label{eq:GravityObserverTerms}
\end{equation}
in the Lagrangian density of the discrete simplicial gravity. In the expression (\ref{eq:GravityObserverTerms}) 
the left-hand side of the equality represents a continuum version
of the theory with the scalar curvature $\sqrt{g}R(x)$ calculated at the 
point of spacetime $x$, while the right-hand side - --a corresponding discretized
version with the sum running over hinges of simplices crossing the point $x$ 
and serving as building blocks of spacetime and $A_h$ being the area of the hinge, $\delta_h$ the associated deficit angle
$\delta_h=2\pi-\sum_{\rm blocks\ meeting\ at\ \theta}\theta$ and $\theta$ is the corresponding dihedral angle.
The field $\phi$ representing von Neumann observers is a source of quenched disorder in the theory which
we consider Gaussian distributed in our simulations.

As was briefly mentioned in the Introduction, since the Regge-Wheeler theory possesses a UV fixed point in the number of dimensions
$D=2,3,4$ \cite{Hamber2000}, the problem  of comparing observables in the presence of disorder (\ref{eq:GravityObserverTerms}) and without it is greatly simplified being reduced to the problem of comparing critical exponents of the theory at the fixed point $k=k_c$.
In particular, we were interested in the dependence
of the universal critical exponent $\nu$ on the
background space dimensionality. As usual, we define the critical exponent $\nu$
through the average space curvature
\begin{equation}
\frac{\langle\int d^{D}x\sqrt{g}R\rangle}{\langle\int
d^{D}x\sqrt{g}\rangle}\sim(k_{c}-k)^{D\nu-1},
\label{eq:DefOfNu}
\end{equation}
where $k=1/8\pi G$ and $k_{c}$ represents the critical point of the theory.

The exponent $\nu$ is directly related to the derivative
of the beta function for the gravitational constant near the ultraviolet
fixed point according to $\beta'(k_{c})=-\nu^{-1}$. Namely, in $D=2+\epsilon$ space dimensions one has (assuming free gravity with a 
cosmological constant) \cite{Weinberg1979,Kawai1990,Aida1995}
\begin{equation}
\frac{1}{8\pi{}k_c}=\frac{3}{50}\epsilon-\frac{9}{250}\epsilon^2+\ldots,
\label{eq:CriticalKc}
\end{equation}
\begin{equation}
\nu^{-1}=-\beta'(k_c)=\epsilon+\frac{3}{5}\epsilon^2+\ldots.
\label{eq:BetaFunction}
\end{equation}

To approach the problem in question, we have performed Monte-Carlo simulations of simplicial Euclidean quantum gravity in $D=4$ space dimensions on hypercubic lattices of sizes $L=4$ ($256$ sites, $3840$ edges, $6144$ simplices), $8$ ($4096$ sites, $6144$ edges, $98304$ simplices) and $16$ ($65536$ sites, $983040$ edges, $1572864$ simplices). In all simulations, the topology was fixed to be the one of $4$-torus, and no fluctuations of topology were allowed. The bare cosmological constant was also fixed to 1 (since the gravitational coupling is setting the overall length scale in the physical problem). To establish efficient thermalization of the system in our numerical experiment (in the absence of disorder) we have investigated behavior of the system at $20$ different values of $k$. For $L=16$ hyper-lattice $33000$ consequent configurations were generated for every single realization of disorder, for $L=8$ hyper-lattice --- $100000$ configurations and for $L=4$ hyper-lattice --- $500000$ consequent configurations. Obtained dependence of the average curvature (\ref{eq:DefOfNu}) was then fit to the singular dependence on $k$ to determine the values of critical gravitational coupling $k_{c}$ and the critical exponent $\nu$.  In the absence of disorder (setting all couplings to the disorder field $\phi_{k}$ to 0) we found for the $L=4$ hyper-lattice $k_{c}=0.067(3)$, $\nu=0.34(5)$, for the $L=8$ hyper-lattice --- $k_{c}=0.062(5)$, $\nu=0.33(6)$ and for the $L=16$ hyper-lattice --- $k_{c}=0.061(7)$, $\nu=0.32(9)$; a relatively weak dependence of the fixed point scale $k_c$ on $L$ pointed out towards efficient thermalization of the employed Euclidean lattice system.

We repeated the same procedure for $10000$ different realizations of the random disorder $\phi_{k}$. Fitting dependence of the average curvature on $k$ for configurations averaged over disorder, we found the value of $k_c$ (post disorder averaging) to be $k_c\approx{}0.03\pm{}0.12$, in principle consistent with $k_c=0$ (compare with $k_c\approx 0.07$ in the case without disorder). We have found that the value $\nu^{-1}=0.01\pm0.06$ for the $L=4$ hyper-lattice, $\nu^{-1}=0.02\pm0.05$ for the $L=8$ hyper-lattice and $\nu^{-1}=0.02\pm0.04$ for the $L=16$ hyper-lattice (compare with $\nu^{-1}\approx3$, which holds approximately in the case without disorder).

In principle, both of these observations (vanishing of $k_c$ and $\nu^{-1}$) --- but especially the second one --- are consistent with a Parisi-Sourlas-like dimensional reduction in the presence of disorder (\ref{eq:GravityObserverTerms}). Indeed, it has been argued previously (see
for example \cite{Hamber2004}) that $\nu\approx\frac{1}{D-1}$ for large $D$, while $\nu=\infty$ exactly for $D=2$. If an analogue of Parisi-Sourlas dimensional reduction holds also for quantum gravity, this naturally implies that the upper critical dimension of
gravity ($D=2$ in the absence of disorder) is lifted to $4$ in the presence of a random network of von Neumann detectors performing measurements of scalar curvature.\footnote{One can naturally ask what happens in simplicial Euclidean quantum gravity (with a quenched disorder) at $D_{\rm cr}>4$ and at $D=3$? If the analogy with behavior of field theories in external fields holds for gravity completely, we expect the theory of simplicial $(D=5)$-dimensional quantum gravity with a quenched disorder to be equivalent to a $3$-dimensional theory without such disorder etc. On the other hand, Parisi-Sourlas correspondence would break down at $D=3$ in a similar fashion as it happens in $D=3$ random field Ising model, see discussion in the Appendix. We leave this question to the future study.} An effective low dimensionality emerging in simulations of simplicial quantum gravity has been previously also reported in \cite{Berg1985,Hamber1985}
where it has been argued that the UV phase of the theory features an effective dimensional reduction with polymer-like behavior of the
correlation functions of observables, while its IR physics is smooth with effectively Euclidean AdS (EAdS) background. Vanishing of the critical value $k_c$ after averaging over quenched disorder (\ref{eq:GravityObserverTerms}) would in turn force one to think that the UV phase becomes the only accessible one across all scales $k$, naively implying unphysical behavior of the theory in the presence of quenched disorder. We shall argue in the next Section that the observed behavior is fully physical and, in sense, a natural one which should be expected from the quantum theory of gravity in the presence of quenched disorder.

Finally, we note in passing that $k_c$ vanishing after averaging over disorder in gravity also seems analogous to a phenomenon which has already been observed in field theories with quenched disorder: for example, the 2nd order phase transition of Ising model (reduced to $\lambda\phi^4$ scalar field theory in the continuum limit) is reached at finite temperature $T_c$ in the absence of disorder and at $T=T_c$ in the presence of random external field
\cite{Fisher1986}.


\begin{figure}[!h]
\centering\includegraphics[width=3.5in]{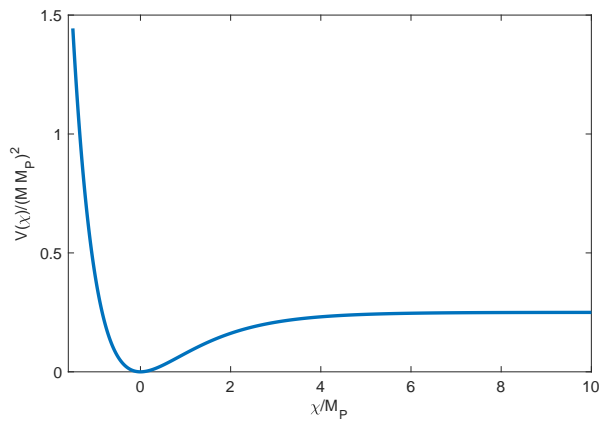}
\caption{
\label{fig:potential}
Effective potential $V(\chi)=\frac{M^{2}M_{P}^{2}}{4}(1-\exp(-\sqrt{2/3}\chi/M_{P}))^{2}$ of the scalar mode $\chi=\sqrt{3/2}M_{P}\log\left(1+\frac{R}{M^{2}}\right)$  related
to spacetime curvature in the Einstein frame. After averaging over quenched disorder, all not-trivial correlations of gravitational degrees of freedom are represented by correlation functions of $\chi$.}
\end{figure}

\section{Physical origin of possible Parisi-Sourlas-like dimensional reduction in Lorenzian
quantum gravity\label{sec:TheorArguments} }
The observed effect of dimensional reduction in Regge-Wheeler simplicial
quantum gravity can be understood (and possibly explained) using the
following theoretical arguments. These arguments also allow to establish
correspondence between the degrees of freedom in the $4-$dimensional
gravity and the the effective $2$-dimensional one.

The continuum limit of the theory (\ref{eq:GravityObserverTerms})
(assuming that it exists) is expected to correspond to a scalar-tensor
Euclidean gravity, where the ``dilaton'' field $\phi$ is sufficiently
massive, so that its arbitrary configuration in the world volume of
the theory can be considered a quenched disorder. If this disorder
is Gaussian, the partition function of the continuum version of the
theory is then given by
\begin{equation}
Z=\int{\cal D}\phi\int\frac{{\cal D}g}{{\cal D}f}\exp\left(-\int d^{4}x\sqrt{g}
\left(\Lambda+(M_{P}^{2}+\phi)R+\frac{1}{2}M^{2}\phi^{2}\right)\right),
\label{eq:PartitionFunctionQG}
\end{equation}
where the integration measure in the path intergral over space metric
$g$ is assumed to be invariant with respect to arbitrary diffeomorphisms
(hence the division by the volume of Diff
group ${\cal D}f$). Integrating over all possible realizations of
$\phi$ disorder, one obtains an effective $f(R)-$theory of gravity
with partition function $Z\sim\int\frac{{\cal D}g}{{\cal D}f}
\exp(-\int d^{4}x\sqrt{g}f(R))$
and
\begin{equation}
f(R)\sim\Lambda+M_{P}^{2}R+\frac{R^{2}}{2M^{2}}.
\label{eq:FofR}
\end{equation}

The version of the same theory obtained by analytic continuation to
spacetimes with Lorentz signature\footnote{The question how such analytic continuation
should be performed technically
is far from trivial; here for the sake of simplicity we shall follow
the naive prescription for the Wick rotation $t\to-it$.} admits de Sitter-like solutions for
all possible values of its parameters
$\Lambda$ and $M$ \cite{Starobinsky1980}, and such solutions represent
dynamical attractors in the phase space of the theory. Indeed, switching
from the Einstein frame to the Jordan frame in the $f(R)-$theory
of gravity, one finds that the theory (\ref{eq:FofR}) is effectively
equivalent to a theory of gravity coupled to a scalar field
\begin{equation}
\chi\sim\sqrt{3/2}M_{P}\log\left(1+\frac{R}{M^{2}}\right),
\label{eq:FieldDef}
\end{equation}
where $R$ is the scalar curvature of spacetime in the original $f(R)$-theory.
As always in analysis of an inflationary theory, we are interested
in the case of super-Planckian $\chi$, meaning that $M^{2}\ll R\ll M_{P}^{2}$
(i.e., the mass scale $M$ is large but well below than the Planckian
scale --- note that this mass is entirely determined by statistical properties of observation events and the coupling strength
between observers and gravitational field).

The potential of this effective scalar field in the Jordan frame is
given by \cite{Barrow,DeFelice2010,BarvinskyKamenshchik1,BarvinskyKamenshchik2,BezrukovShaposhnikov} 
\begin{equation}
V(\chi)\sim\frac{M^{2}M_{P}^{2}}{4}(1-\exp(-\sqrt{2/3}\chi/M_{P}))^{2},
\label{eq:Potential}
\end{equation}
which reduces to the potential of chaotic inflation at small $\chi\ll M_{P}$
and a potential quickly approaching a constant asymptotics at $\chi\gg M_{P}$.
We are primarily interested in the regime, where
$\chi>0$ and large, which according to (\ref{eq:FieldDef}) corresponds
to the positive scalar curvature of the spacetime in the Einstein
frame. However, nothing prevents us from considering the case $\chi<0$,
$|\chi|\gg M_{P}$ as well which again corresponds to the slow roll inflation
in the Jordan frame, while describing anti-de Sitter physics in the
Einstein frame (with $R$ bounded from below by the parameter $M^{2}$,
again interestingly depending on the statistical properties of the
distribution of observers and observation events in the spacetime).

Returning to the case in question with $\chi>0$ and integrating out
sub-horizon fluctuations of the effective field $\chi$ (such fluctuations
can be considered Gaussian in the first approximation due to applicability
of EFT approximation for gravitational degrees of freedom in the UV), one arrives
to the physical picture of an inflationary self-reproducing universe
with the only survived ``coordinates'' being the number of inflationary
efoldings (log of scale factor) and an effective scalar field $\chi$
(essentially, a log of scalar curvature in the Einstein frame); in
this sense, the originally $(3+1)$-dimensional theory becomes effectively $2$-dimensional in the infrared. Let us
show in details that this is indeed the case using stochastic inflationary
formalism \cite{Starobinsky1988,Sasaki1988,Nambu1989,Starobinsky1994,Rey1987,Hosoya1989,Graziani1988,Lawrie1989,Podolsky2002,Enqvist2008}
and ignoring gravitational vector and tensor modes which do not contribute
to quasi-de Sitter gravitational entropy \cite{Podolskiy2018}  and thus do not influence strongly infrared dynamics of the theory.

Namely, separating the field $\chi$ into the subhorizon and superhorizon
parts, one can write:
\begin{equation}
\chi(t,x)=\chi_{IR}(t,x)+\frac{1}{(2\pi)^{3/2}}\int d^{3}k\cdot\theta(k-\epsilon aH)
\left(a_{k}\phi_{k}(t)\exp(-ikx)+{\rm h.c.}\right)+\delta\chi,
\label{eq:IR-UV}
\end{equation}
where $a(t)$ is the scale factor of de Sitter spacetime, $H$ is
the corresponding Hubble constant, $\theta(...)$ is the Heaviside
step-function, the modes $\phi_{k}(t)=\frac{H}{\sqrt{2k}}(\tau-\frac{i}{k})\exp(ik\tau)$
correspond to the Bunch-Davies de Sitter invariant vacuum of a free
massless scalar field, $\tau=\int\frac{dt}{a(t)}$, $\epsilon$ is
a small number such that $\epsilon\ll1$ (which determines a notation
for separating superhorizon modes from the subhorizon ones) and $\delta\chi$
can be neglected in the leading order with respect to $H/M_{P}$.
Substituting this decomposition into the equation of motion for the
field $\chi$ on de quasi-Sitter background, one obtains the equation
for the infrared part of the field $\chi_{IR}$:
\begin{equation}
\frac{d\chi_{IR}}{d\tau}=-\frac{1}{3H^{2}}\frac{dV}{d\chi_{IR}}+\frac{f(\tau,x)}{H},
\label{eq:Langevin}
\end{equation}
where a composite operator
\[
f(\tau,x)=\frac{\epsilon aH^{2}}{(2\pi)^{3/2}}\int d^{3}k\cdot\delta(k-\epsilon aH)
\cdot\frac{(-i)H}{\sqrt{2}k^{3/2}}\left[a_{k}\exp(-ikx)-a_{k}^{\dagger}\exp(ikx)\right]
\]
has the correlation properties
\[
\langle f(\tau,x)f(\tau',x)\rangle=\frac{H^{4}}{4\pi^{2}}\delta(\tau-\tau'),
\]
if the average is taken over the Bunch-Davies vacuum state. Another
very important property of this operator is that its self-commutator
vanishes, and thus the equation (\ref{eq:Langevin}) can be considered
a stochastic differential equation for the quasi-classical but stochastically
distributed long-wavelength field $\chi_{IR}$ (from now on, we shall
drop the index $IR$ always implying that the infrared, superhorizon
part of the field $\chi$ is considered).

One then obtains an effective Fokker-Planck
equation (see for example Appendix \ref{sec:SFP}) corresponding to the Langevin equation (\ref{eq:Langevin})
for the probability $P(\tau,\chi)$ to measure a given value of the
background/infrared scalar field $\chi$ in a given Hubble patch:
\begin{equation}
\frac{\partial P}{\partial\tau}\approx\frac{1}{3\pi M_{P}^{2}}
\frac{\partial^{2}}{\partial\chi^{2}}(VP)+
\frac{M_{P}^{2}}{8\pi}\frac{\partial}{\partial\chi}
\left(\frac1V\,\frac{dV}{d\chi}P\right),
\label{eq:FP}
\end{equation}
where $\approx$ implies that the equation (\ref{eq:FP}) by itself
is an approximation (we made a number of simplifications during its
derivation such as neglecting subdominant terms $\delta\chi$ in the
expansion (\ref{eq:IR-UV}), assuming slow roll of the field $\chi$
and neglecting self-interaction of the field $\chi$ at subhorizon
scales). We thus assume that it holds on average and only approximately,
and model it by including an additional term $F(\tau,\chi)$ to its
right-hand side, again quasi-classical but stochastic (see the next Section). Taking into
account the smallness of this term, assuming its Gaussianity (so that
$\langle F(\tau,\chi)F(\tau',\chi')=\Delta\delta(\tau-\tau')\delta(\chi-\chi')$)
and integrating it out, we finally conclude that the infrared dynamics
of the theory (\ref{eq:PartitionFunctionQG}) is being essentially
determined by the partition function
\[
Z_{IR}=\int{\cal D}P\exp(-{\cal W}),
\]
where the effective action ${\cal W}$ of the theory is given by
\begin{equation}
{\cal W}=\int d\tau d\chi\frac{1}{\Delta}\left(-\frac{\partial P}{\partial\tau}+
\frac{1}{3\pi M_{P}^{2}}\frac{\partial^{2}}{\partial\chi^{2}}(VP)+
\frac{M_{P}^{2}}{8\pi}\frac{\partial}{\partial\chi}\left(\frac{dV}{Vd\chi}P\right)\right)^{2}.
\label{eq:QuadraticEffAction}
\end{equation}
It is now instructive to use the de Sitter ``entropy'' $S$ defined
according to the prescription $P(\tau,\chi)=\exp(S(\tau,\chi))$ instead
of the probability distribution $P$. One motivation for this substitution
is the fact that the distribution function $P(\tau,x)$ converges
to
\[
P(\tau\to\infty,\chi)\sim\frac{1}{V(\chi)}\exp\left(\frac{3M_{P}^{4}}{8V(\chi)}\right)
\]
 in the limit $\tau\to\infty$, where the expression in the exponent
coincides exactly with the gravitational entropy of de Sitter space.
As we shall see below, there are other advantages of using $S$ instead
of $P$.

One finds after the substitution
\[
{\cal W}=\int d\tau d\chi\left(\frac{e^{2S}}{\Delta}\left[-\frac{\partial S}{\partial\tau}+
\frac{V}{3\pi M_{P}^{2}}(S''+(S')^{2})+(\frac{2V'}{3\pi M_{P}^{2}}
+\frac{M_{P}^{2}}{8\pi}(\log V)')S'+...\right.\right.
\]
\begin{equation}
\left.\left.(\frac{V''}{2\pi M_{P}^{2}}+\frac{M_{P}^{2}(\log V)''}{8\pi})\right]^{2}-S\right),
\label{eq:Seffaction}
\end{equation}
where prime denotes partial differentiation with respect to the field
$\chi$, and the appearance of the last term is due to the Jacobian
in the measure of functional integration emerging after the change
of functional variables.

Substituting the particular form of the potential (\ref{eq:Potential})
of interest for us to the expression (\ref{eq:Seffaction}), we obtain
\begin{equation}
{\cal W}=\int d\tau d\chi\left[\frac{e^{2S}}{\Delta}\left(-\frac{\partial S}{\partial\tau}+
\frac{M_{P}}{4\pi}\sqrt{\frac{2}{3}}z\frac{\partial S}{\partial\chi}
-\frac{z}{6\pi}\right)^{2}+\ldots\right],
\label{eq:QuasiLiouvilleAction}
\end{equation}
where $z=\exp\left(-\sqrt{\frac{2}{3}}\frac{\chi}{M_{P}}\right)$.
In the quasi-de Sitter limit $S=S_{0}+\delta S$, $\delta S\ll S_{0}$,
the potential term in this action coincides with the one of a Liouville-like
theory of the ``field'' $S$ in a two-dimensional spacetime spanned
by the coordinates $(\tau,\chi)$, i.e., the two-dimensional theory
of quantum gravity \cite{Polyakov1981,Knizhnik1988} with the ``field''
$S$ playing the role of the conformal mode of the 2-dimensional spacetime metric.

In other words, in the absence of anisotropic stress covariant observables
in quantum gravity can be expressed in terms of correlation functions
of the scalar curvature (in one-to-one correspondence with the scalar
degree of freedom $\chi$ in the Einstein frame) according to the
prescription
\begin{equation}
\langle\chi^{n}\rangle\text{\ensuremath{\sim}}
\langle(M_P\log\left(\frac{R}{M^2}\right))^{n}\rangle\text{\ensuremath{\sim}}
\int d\chi\cdot\chi^{n}\int{\cal D}S\exp(-{\cal W}),
\label{eq:OnePointCorrelationFunctions}
\end{equation}
where the effective action in the path integration (\ref{eq:OnePointCorrelationFunctions})
is given by the expression (\ref{eq:QuasiLiouvilleAction}). Therefore,
the limit $\log a\to\infty$ of the integrand in Eq. (\ref{eq:OnePointCorrelationFunctions})
can be thought of as a ground state of the theory of gravity (\ref{eq:FofR}).

To formalize the map (\ref{eq:OnePointCorrelationFunctions}) a bit clearer, we can write averages of any observable $\langle{\cal O}\rangle{}(t)$
at the time $t\sim H^{-1}\log (a)$ as
\begin{eqnarray}
    \langle {\cal O}(\chi)\rangle(t) =\int d\chi\,{\cal O}(\chi)\,P(\chi,t),
\end{eqnarray}
where the partition function $P$ satisfies the Fokker-Planck equation 
\begin{eqnarray}
    E(\dot P,\partial_\chi P,\chi)\equiv -\dot P+\hat{H}P=0.
\end{eqnarray}
We thus have a chain of transformations
\begin{eqnarray}
    \langle {\cal O}(\chi)\rangle(t) =\int d\chi\,{\cal O}(\chi)\,P(\chi,t)
    =\int d\chi\,{\cal O}(\chi)\,\int D\bar P\,\bar P(\chi,t)\,\delta[\, E(\dot{\bar P},\partial_\chi \bar P,\chi)\,],
\end{eqnarray}
where the Jacobian for transformation between $E$ and $P$ is disregarded for simplicity. The functional delta-function in the last
integral on the right is effectively regulated by the small parameter $\Delta\to0$ according to 
\begin{eqnarray}
    &&\delta[\, E(\dot{\bar P},\nabla \bar P,\chi)\,]=\prod\limits_{t,\chi}
    \delta\Big(E\big(\dot{\bar P}(t,\chi),\partial_\chi \bar P(t,\chi),\chi\big)\Big)\equiv
     \prod\limits_x\delta\Big(E\big(\nabla{\bar P}(x),\nabla \bar P(x),\chi\big)\Big)\nonumber\\
     &&\qquad\qquad\qquad\qquad=\exp\Big(-\frac1\Delta\int d^2x\,
     E^2\big(\nabla{\bar P}(x),\nabla \bar P(x),\chi\big)\Big).
\end{eqnarray} 
Here we introduced 2D coordinates $(x_0,x_1)=(\log(a),\chi)$, and $\nabla=\partial_0,\partial_1$ collects all 2D derivatives. Thus we have 
(dropping bar over functional integration variable $P$)
\begin{eqnarray}
    &&\langle {\cal O}(\chi)\rangle(t) =\int D\bar P\,\exp\Big(-\frac1\Delta\int d^2x\,
     E^2\big(\nabla{\bar P}(x),\nabla \bar P(x),\chi\big)\Big)\,\int d\chi\,{\cal O}(\chi)\,\bar{P}(\chi,t)\nonumber\\
     &&\qquad\qquad=\int D P\,\exp\Big(-\frac1\Delta\int d^2x\,
     E^2\big(\nabla{ P}(x),\nabla P(x),\chi\big)\Big)\,\int d^2y\,{\cal O}(y^1)\,P(y)\,\delta(y^0-t).\nonumber
\end{eqnarray} 
The probability $P$ to measure a given value of $\chi$ in a given Hubble patch is then re-parametrized as $P\sim\exp(S)$, and $S$ become the field
variable of interest for us. (Note that in the quasi-de Sitter limit $S$ coincides with the gravitational entropy of de Sitter space in the limit $\log(a)\to 0$.) The mapping dictionary of duality between 2D side and 4D side for the action in quantum measure and for observables of interest is then defined
according to the prescription
\begin{eqnarray}
    &&\exp\Big(-\frac1\Delta\int d^2x\,
     E^2\big(\nabla{\bar P}(x),\nabla \bar P(x),\chi\big)\Big)
     \quad\Leftrightarrow\quad
     \exp\Big(-{\rm{}Polyakov\; action}\Big)      \label{eq:mapping}\\
     &&{\cal O}(\chi)\quad\Leftrightarrow\quad\int d^2y\,{\cal O}(y^1)\,P(y)\,\delta(y^0-t),
 \end{eqnarray}

The physical reason why the dimensional reduction has effectively
realized in the theory (\ref{eq:FofR}) and its analytic continuation
to Lorentzian spacetimes is simple: once the dynamics of relevant
degrees of freedom is coarse-grained to comoving spatio-temporal scales
$\sim H_{0}^{-1}\sim\Lambda^{-1/2}$ (as we are interested in the
continuum limit of the theory (\ref{eq:GravityObserverTerms}), it
is natural to study exactly this case), the global structure of spacetime
is represented by a set of causally unconnected Hubble patches; expectation
values and correlation functions of the field $\chi$ are determined
by a stochastic process generated by the Langevin equation (\ref{eq:Langevin}),
values of $\chi$ in different Hubble patches are completely independent
of each other, and thus the spatial dependence of $\chi$ becomes
largely irrelevant.

We have argued that the $4$-dimensional gravity with a quenched ``dilaton''
$\chi$ becomes reduced to an effectively two-dimensional theory in
the deep infrared limit (of large spatio-temporal coarse-graining), where
coordinate mapping of the fluctuating spacetime is given in terms
of the number of efoldings $\tau=\log a$ and the effective scalar
degree of freedom $\chi$ related to the large-scale curvature of
spacetime in the Einstein frame according to the prescription (\ref{eq:FieldDef}).
We emphasize that the physical scales at which this description becomes efficient coincide and exceed the scales of eternal inflation 
from the point of view of a subhorizon observer, thus effectively regularizing the structure of the theory in this deep IR limit.
Tensor and vector degrees of freedom present in the metric for the subhorizon observer are effectively integrated away and do not contribute to the infrared structure of the correlation functions of observables in the theory. When the probe scale approaches the cosmological horizon scale, this effectively 2D physics has to be matched to an effective 4D field theory description of gravitational degrees of freedom, and it is quite clear from the setup how it has to be done physically (effective subhorizon 4D degrees of freedom including vector and tensor ones are propagating on the stochastic background with large scale statistical properties effectively determined by the Liouville physics described above). 

\section{Fokker-Planck equation and its extensions in the two-noise model\label{sec:TwoNoiseSec}}

In this Section, we shall derive the effective action (\ref{eq:QuadraticEffAction}) used above, albeit in a schematic fashion, and estimate
dependence of the parameter $\Delta$ in (\ref{eq:QuadraticEffAction}) on $\delta$ and slow roll parameters. 

As was discussed previously, the inflationary 
Fokker-Planck equation holds its canonical celebrated form (\ref{eq:FP}) only in the regime $\delta\to0$, $\epsilon_H\to0$, which does not necessarily hold anywhere except very close to the de Sitter spacetime geometry. Moreover, even for geometries globally close to $dS_4$ one might be interested in behavior of the IR effective theory under different values of parameter $\delta$ separating IR and UV physics (at this point, one would only be aware of the fact the the
theory approaches the regime $\Delta\to{}0$ with Fokker-Planck-like dynamics of $P(\chi,N)$ at $\delta\to{}0$, which is entirely
independent of $\delta$). In short, we would like to derive extension of this equation which would hold to first order in slow roll parameters $\epsilon_H,\, \eta_H$ and, ideally, to order in $\delta$ higher than first.

First of all, one notes that the one-noise stochastic model for the infrared dynamics of the scalar field in the inflationary
spacetime (\ref{eq:Langevin})-(\ref{eq:FP}) cannot be used for this derivation as it produces manifestly non-local results, see Appendix \ref{sec:NonLocal}. This non-locality stems from the presence of additional degree of freedom which is integrated out to obtain the effective theory (\ref{eq:Langevin})-(\ref{eq:FP}) in the case of generic $\epsilon_H,\,\delta$. It can be shown that this degree of freedom can be accounted for if we consider a two-noise model similar to the one introduced in \cite{Nambu1989}:
\begin{equation}
\frac{d\Phi}{dN}=\frac{v}{H}+\sigma,
\label{eq:2NoiseLangevin1}
\end{equation}
\begin{equation}
\frac{dv}{dN}=-3v-H^{-1}\frac{\partial V}{\partial \Phi}+\tau,
\label{eq:2NoiseLangevin2}
\end{equation}
where
\begin{equation}
\sigma(N,{\bf x})=\frac{1}{H}\int \frac{d^3k}{(2\pi)^{3/2}}\delta(k-\delta{}aH)\left(a_k\phi_k e^{-i\textbf{kx}}+{\rm h.c.}\right)
\label{eq:sigmadef}
\end{equation}
\begin{equation}
\tau(N,{\bf x})=\frac{1}{H}\int \frac{d^3k}{(2\pi)^{3/2}}\delta(k-\delta{}aH)\left(a_k\dot{\phi}_k e^{-i\textbf{kx}}+{\rm h.c.}\right)
\label{eq:taudef}
\end{equation}
The expressions for the modes $\phi_k$, $\dot{\phi}_k$ have to be derived under assumption of finite (but small $\epsilon_H$). Importantly, to the first order in small roll parameters we can keep $\epsilon_H$ constant (see Appendix \ref{sec:NonLocal}). The equation for the modes $u_k=a\phi_k$ then has the form
\begin{equation}
0=u_k''+\left(k^2-\frac{a''}{a}+m^2a^2\right)u_k=u_k''+\left(k^2-\left(2-\epsilon_H-\frac{m^2}{H^2}\right)H^2a^2\right),
\end{equation}
where $m$ is the effective mass of the scalar field. To the leading order, we have $\epsilon_H\approx{}\frac{m^2}{3H^2}$, and thus $2-\epsilon_H-\frac{m^2}{H^2}\approx{}2-4\epsilon_H=2(1-\epsilon_H)$. One the other hand, again, $a\approx-\frac{1}{H\eta}(1+\epsilon_H)+{\cal O}(\epsilon_H^2)$, and we finally obtain that (to the leading linear order in slow roll parameters $\epsilon_H$) the field $u_k$ satisfies the free massless field equation
\begin{equation}
u_k''+\left(k^2-\frac{2}{H^2\eta^2}\right)=0,
\end{equation}
with its properly normalized solution given by
\begin{equation}
u_k(\eta)=-\frac{1}{\sqrt{2k}}\left(1-\frac{i}{k\eta}\right)\exp(-ik\eta)+{\cal O}(\epsilon_H^2).
\end{equation}
Thus, to the first order in $\epsilon_H$ the mode $\phi_k$ is given by
\begin{equation}
\phi_k(\eta)=\frac{H(1+\epsilon_H)}{\sqrt{2k}}\left(\eta-\frac{i}{k}\right)\exp(-ik\eta)+{\cal O}(\epsilon_H^2)
\label{eq:ModeChi}
\end{equation}
Differentiating this expression with respect to world time $t=\int\frac{dN}{H}$ we find
\begin{equation}
\frac{d\phi_k(\eta)}{dt}=\frac{ikH^2\eta^2}{\sqrt{2k}}e^{-ik\eta}
-\frac{\epsilon_HH^2}{\sqrt{2k}}\left(\eta-\frac{i}{k}\right)e^{-ik\eta}+{\cal O}(\epsilon_H^2).
\label{eq:ModeChiDerivative}
\end{equation}

The correlation functions of the noise terms $\sigma(N)$ and $\tau(N)$ (as usual, we are interested only in the behavior of correlation functions 
in the same Hubble patch parametrized by the same ``coarse-grained'' spatial point $\textbf{x}$) are in turn found to be
\begin{equation}
\langle\sigma(N)\sigma(N')\rangle\approx\frac{H^2}{4\pi^2}\delta(N-N')\left(1+3\epsilon_H+\delta^2\right),
\label{eq:SigmaSigma}
\end{equation}
\begin{equation}
\langle\tau(N)\tau(N')\rangle\approx\frac{\delta^2H^4}{4\pi^2}\delta(N-N')\left(\delta^2+2\epsilon_H\right)
\label{eq:TauTau}
\end{equation}
\begin{equation}
\langle\tau(N)\sigma(N')\rangle\approx\delta(N-N')\frac{H^3}{4\pi^2}(\epsilon_H+\delta^2-i\epsilon_H\delta)
\label{eq:TauSigma}
\end{equation}
(note that the latter two correlation functions vanish in the limit $\delta\to0$), and the correlation function $\langle\sigma(N)\tau(N')\rangle$ is related to (\ref{eq:TauSigma}) by complex conjugation. Mixed $\tau\sigma$ correlators also end up suppressed by either powers of $\delta$ or $\epsilon_H$. Note in this respect that one has to be careful taking one of the limits $\delta\to0$ or $\epsilon_H\to0$ first (compare for example to \cite{Nambu1989}). Two limiting cases are of special interest:\\
\textbf{(a) Quasi-de Sitter limit.} $\epsilon_H\ll\delta^2<1$, the case considered in the Appendix \ref{sec:NonLocal} with $\epsilon_H$ negligible but keeping all orders in $\delta$
\[
\langle\sigma(N)\sigma(N')\rangle\approx\frac{H^2}{4\pi^2}(1+\delta^2)\delta{}(N-N'),
\]
\[
\langle\tau(N)\tau(N')\rangle\approx\frac{\delta^4H^4}{4\pi^2}\delta(N-N')
\]
\[
\langle\tau(N)\sigma(N')\approx\frac{\delta^2H^3(1+i\delta)}{4\pi^2}\delta(N-N')	
\]
and\\
\textbf{(b) ``Deep IR physics'' or Nambu-Sasaki limit.} $\delta^2\ll\epsilon_H\ll1$:
\[
\langle\sigma(N)\sigma(N')\rangle\approx\frac{H^2}{4\pi^2}(1+3\epsilon_H)\delta{}(N-N'),
\]
\[
\langle\tau(N)\tau(N')\rangle\approx\frac{2\delta^2\epsilon_HH^4}{4\pi^2}\delta(N-N')\approx{}0,	
\]
\[
\langle\tau(N)\sigma(N')\rangle\approx\frac{\epsilon_HH^3}{4\pi^2}\delta(N-N')	
\]
While both cases are very illustrative and somewhat similar (specifically, in the regime $\delta\ll1$), here for our purposes we will focus on the first one, in which functional integrations are simplified greatly. The opposite case (b) is considered in relative depths in \cite{Nambu1989} and will be discussed in more details in a subsequent work.

To derive the Fokker-Planck equation and corrections to it, we follow the path integral approach outlined in the Appendix \ref{sec:SFP}. It can be seen easily that the diffusion matrix associated with the correlation properties of the noises is singular in the quasi-de Sitter limit (a), and the functional integration measure for the noise terms has the form
\begin{equation}
Z_{\rm noise}=\int{\cal D}\sigma{\cal D}\tau\exp\left(-\frac{1}{2}\int{}dN\, f^T{}D^{-1}f \right),
\label{eq:NoisePartitionFunction}
\end{equation}
where 
\begin{equation}
D=
\begin{pmatrix}
\frac{H^2(1+\delta^2)}{4\pi^2} & \frac{H^3\delta^2(1+i\delta)}{4\pi^2} \\
\frac{H^3\delta^2(1-i\delta)}{4\pi^2} & \frac{\delta^4H^4}{4\pi^2} 
\end{pmatrix}
\label{eq:DiffCoeffMatrix}
\end{equation}
and $f^T=(\sigma,\tau)$. The matrix $D$ is manifestly singular, and the noises $\sigma$ and $\delta$ are correlated. Calculating eigenvectors and eigenvalues of the matrix $D$, we find that
\begin{equation}
\tau=-H\sigma\frac{i\delta^2}{i+\delta},
\label{eq:NoiseCorrelation}
\end{equation}
while the non-trivial contribution into (\ref{eq:NoisePartitionFunction}) is given by the combination $H^{-1}\tau+\frac{i(\delta-i)}{\delta^2}\sigma$.
Since the matrix $D$ is singular only in the limit of vanishing slow roll parameters $\epsilon_H\to0$, it should be kept in mind that its eigenvector corresponding to the zero eigenvalue really introduces a constraint on the dynamics of $v$ and $\Phi$.

The correlation functions of $\Phi$ and $v$ can in turn be obtained by integrating over the measure
\[
F=\int{\cal D}\sigma{\cal D}\tau{\cal D}v{\cal D}\Phi{\cal D}\lambda{\cal D}\mu\exp\left(\int{}dN\left(
i\lambda\left(\frac{\partial{}v}{\partial{}N}+3v+\frac{\partial{}V/\partial\Phi}{H}-\tau\right)+\right.\right.	
\]
\begin{equation}
+\left.\left.i\mu\left(\frac{\partial\Phi}{\partial{}N}-\frac{v}{H}-\sigma\right)\right)\right)Z_{\rm noise}.	
\label{eq:Measure}
\end{equation}
As all integrations (with the exception of the integration over $\Phi$) are Gaussian, 
they can be explicitly taken revealing
\[
	F=\int{\cal D}\Phi{\cal D}v\exp(-S),
\]
where
\[
S=\int dN{\cal L}=\frac{1}{2}\int{}dN\,\frac{2\pi^2}{H^2(1+\delta^2+\delta^4)}\left(\frac{1+i\delta}{\delta^2}\left(
\frac{\partial\Phi}{\partial{}N}-\frac{v}{H}\right)+\right.
\]
\begin{equation}
+\left.\frac{1}{H}\left(\frac{\partial{v}}{\partial{}N}+3v+\frac{1}{H}\frac{\partial{}V}{\partial\Phi}\right)\right)^2.
\label{eq:TwoNoiseEffectiveAction}
\end{equation}
On top of this effective action there is a constraint present in the system (the one corresponding to the vanishing eigenvalue of the matrix (\ref{eq:DiffCoeffMatrix})):
\begin{equation}
\frac{\delta^2}{1-i\delta}\left(\frac{\partial\Phi}{\partial{}N}-\frac{v}{H}\right)=\frac{1}{H}
\left(\frac{\partial{}v}{\partial{}N}+3v+\frac{1}{H}\frac{\partial{}V}{\partial\Phi}\right)
\label{eq:Constraint}
\end{equation}
Solving it for $\frac{\partial\Phi}{\partial{}N}-\frac{v}{H}$ and substituting back into the action (\ref{eq:TwoNoiseEffectiveAction}), we obtain the effective theory of the field $v$:
\begin{equation}
Z=\int{\cal D}v{\cal D}\Phi\,\exp\left(-\frac{2\pi^2K}{H^4}\left(\frac{\partial{}v}{\partial{}N}+3v+
\frac{1}{H}\frac{\partial{}V}{\partial{}\Phi}\right)^2\right),
\label{eq:EffTheoryV}
\end{equation}
where $K=(1+\delta^2)^2/(\delta^8\cdot(1+\delta^2+\delta^4))$ (in this representation, $\Phi$ is considered an external field which we average out). 

The conjugate momentum for the field $v$ is given by
\begin{equation}
p_v=\frac{\pi^2K}{H^4}\left(\frac{\partial{}v}{\partial{}N}+3v+\frac{1}{H}\frac{\partial{}V}{\partial{}\Phi}\right)
\label{eq:VConjugateMomentum}
\end{equation}
and the Hamiltonian of the theory (\ref{eq:EffTheoryV}) is
\begin{equation}
{\cal H}_v=-\frac{H^4}{\pi^2K}p_v^2-3v-\frac{1}{H}\frac{\partial{}V}{\partial\Phi}.
\label{eq:HamiltonianV}
\end{equation}
The Fokker-Planck equation for the probability $P(v,N)$ to measure a given value of $v$ in a given Hubble patch is then obtained by writing
down $\frac{\partial{}P(v,N)}{\partial{}N}={\cal H}_v(p_v,v)P(v,N)$ and promoting conjugate momentum $p_v$ into a differential operator according to the usual prescription $p_v=-\partial_v$, see Appendix \ref{sec:SFP}.

An important conclusion is that the theory with the Hamiltonian (\ref{eq:HamiltonianV}) is generally unstable, with a run-away behavior of the probability $P(v,N)$. It can be shown that this conclusion survives in the general case, independent on relations between $\epsilon_H$ and $\delta$ (as we shall demonstrate in the follow-up work): namely, the run-away behavior is associated with the behavior of the probability distribution $P$ as a function of $p_v$, while its behavior as a function of $p_\Phi$ and $\Phi$ remains stable. This is a reflection of general instability of de Sitter space \cite{Polyakov:2012uc}. One and perhaps the only way to deal with this instability is to set a general constraint $p_v=0$ (i.e., to choose initial conditions for the physical system in a rather special way). Then, from the constraint (\ref{eq:Constraint}) we obtain
\begin{equation}
v=H\frac{\partial\Phi}{\partial{}N}
\end{equation}
and substituting it back into the effective action of the theory (\ref{eq:TwoNoiseEffectiveAction}), we finally obtain the theory with Langrangian
\begin{equation}
{\cal L}=\frac{2\pi^2}{H^4(1+\delta^2+\delta^4)}\left(\frac{\partial}{\partial{}N}\left(H\frac{\partial\Phi}{\partial{N}}\right)+
3H\frac{\partial{}\Phi}{\partial{}N}+\frac{1}{H}\frac{\partial{}V}{\partial\Phi}\right)^2.
\label{eq:LangrangianFinal}
\end{equation}
It is then straightforward to show that the theory (\ref{eq:LangrangianFinal}) produces the canonical Starobinsky-Fokker-Planck equation with a singular correction $\sim{}\Delta{}\delta(\Phi-\Phi')$ originating from the first term in parentheses in (\ref{eq:LangrangianFinal}) and $\Delta\sim{}H^2$.

\section{Conclusion\label{sec:Conclusion}}
Numerical and theoretical analysis of non-renormalizable field theories and 4-dimensional quantum gravity
performed here shows that introducing a network of von Neumann observers distributed
in the world volume of the theory and continuously measuring the gravitational field
strength (or scalar curvature in the case of gravity) leads to a drastic non-perturbative restructuring of the Hilbert
space of the underlying theory significantly changing its infrared structure. Perhaps, most importantly, integrating out observers 
induces a de Sitter-like background of the theory which completely determines the infrared structure of correlation functions of its physical observables.
The induced cosmological constant is determined by the properties of observers --- the distribution of observation events in the world volume of the theory and coupling strength between observers and gravitational degrees of freedom. On the one hand, restructuring of the Hilbert space of the gravity coupled to observers is similar to the phenomenon of Anderson-like localization in disordered media. On the other hand, it is characterized by an effective dimensional
reduction close in spirit to the celebrated Parisi-Sourlas dimensional reduction
observed in several field theories in the presence of random external field (such as continuum limit of RFIM --- random field Ising model). 

In the case of gravity, this Hilbert space restructuring can be roughly characterized as follows.
It is known by now that 4D simplicial Euclidean quantum gravity admits a UV fixed point at a particular value of $G_c=1/(8\pi{}k_c)$, with a 
UV, strongly coupled phase at $k<k_c$ and an IR, weakly coupled phase, which is realized at $k>k_c$. The strongly coupled phase admits
an EAdS like behavior of the ground state of gravity and a non-trivial infrared dynamics of correlation functions of observables. 
On the other hand, the weakly coupled phase (which should be the one of physical interest as the real world gravity is weakly coupled!)
seems to feature a quasi-2-dimensional branching polymer-like behavior without any smooth background geometry in the IR. This was previously 
interpreted as an absence of a proper continuum limit of the theory in this regime. Instead, we believe that this behavior is actually physical:
in the weakly coupled regime the ground state of the theory admits a dS-like physics, and a 2-dimensional branching polymer, self-reproducing 
behavior of observables is really nothing but a Wick-rotated equivalent of the eternal inflation happening on this dS-like background. 

Quenched disorder associated with random networks of observers measuring the strength of gravitational interaction clears up the mist 
somewhat in this respect: it seems to move the critical point of the theory towards $k_c\to{}0$ implying that the only accessible phase of
the theory is the one of weakly coupled gravity. Our theoretical analysis further shows that the nature of the effective dimensional reduction
($4D\to 2D$) in the presence of quenched disorder is associated with with the fact that infrared dynamics of observables in the eternally
inflating Universe is determined by the probability $P(\chi,\log (a))=\exp{}(S(\chi,\log(a)))$ to measure a given value of the effective 
``inflaton'' $\chi$ in a given Hubble patch (in other words, all correlation functions of physical observables are entirely determined by
the structure of $P(\chi,\log (a))$). We find this finding rather interesting.

The phenomenon observed here might also explain what should exactly be understood by the continuum
limit of quantum gravity. Indeed, it is generally accepted that the formal continuum limit of non-renormalizable
quantum field theories (including in principle $4$-dimensional quantum general relativity, which might be non-perturbatively
renormalizable) does not exist. Nevertheless, it is still possible to make a number of conclusions regarding the physical properties
of such theories in the large-scale/infrared limit. To a degree, the way how to do it can be understood
 using the correspondence between relativistic quantum field theories and the corresponding statistical classical
field theories obtained from the former by Wick rotation \cite{Itzykson1991}. According to this correspondence, the classical statistical counterpart
of a renormalizable quantum field theory describes behavior of the
order parameter of a classical statistical system in a vicinity of
a second order phase transition. At temperatures close to $T_{c}$
the correlation length $\xi\sim|T-T_{c}|^{-\alpha}$ of the relevant
degrees of freedom approaches infinity, which makes it possible to
describe correlation functions of observables in terms of a small
number of continuous order parameters only. Similarly, the statistical
physics counterpart of a \emph{non-renormalizable} quantum field theory describes
a vicinity of \emph{a first order} phase transition, when the correlation
length $\xi$ of physical degrees of freedom remains finite at all
accessible values of thermodynamic potentials (which forces one to
conclude that the continuum limit of corresponding quantum field theories
does not exist). The process of measuring the physical state of
the field in an equivalent QFT can be thought of as an insertion of
a projection operator in the world volume of the theory at a point
of spacetime, where and when the measurement/observation is performed.
In the statistical mechanical counterpart, such insertion is akin
to an introduction of a heavy, ``quenched'' impurity in the spatial volume of the
classical thermodynamic system, with elementary excitations of the
order parameter(s) scattering against it. It can thus be expected
that the network of von Neumann observers in a QFT is reminiscent
of an ensemble of impurities introduced into the classical system
described by a statistical counterpart of the theory.

It is well known that in the vicinity of a first order phase transition,
such impurities serve as nucleation centers for bubbles of the true
phase \cite{Slezov2009}. When coupling constants of the von Neumann
detectors to the field are sufficiently large, bubble nucleation
process proceeds ad infinitum in a quasi-continuum limit $T\to T_{c}$.
Correspondingly, in a QFT, the vacuum state remains largely inhomogeneous
even in the limit of Langevin time $\tau\to\infty$; the resulting
state is also strongly dependent on the particular location of inserted
operators describing observation events. Thus, the structure of the
Hilbert space of a non-renormalizable quantum field theory is largely
determined by ``localization properties'' of the effective potential
of impurities inserted into an equivalent statistical mechanical system.

To conclude, two observations presented here point out towards a possible $4D\to{}2D$ dimensional reduction
of quantum gravity (at least in the infrared) in the presence of random networks of observers: (a) lattice simulations of simplicial 
Regge-Wheeler Euclidean gravity in the presence of disorder showing that after averaging over such
disorder the critical exponent(s) of the theory change as if the effective
dimensionality of the theory changes from $D=4$ to $D=2$, and (b)
theoretical analysis of quantum (Lorenzian) general relativity in
the presence of quenched disorder, which also hints towards effective dimensional
reduction and also allows to identify relevant degrees of freedom
in the reduced, effectively two-dimensional, theory. Although these
two separate approaches lead to the same conclusion regarding the
physical system in question, ultimately only numerical simulations
Lorentzian quantum gravity will allow to reconcile the two approaches.
We hope to return to this subject in our future studies.
\appendix
\section{Parisi-Sourlas dimensional reduction in non-renormalizable field
theories in the presence of observer networks\label{sec:Parisi-Sourlas-dr}}
Quantum gravity can be thought of as a ``quantum field theory'' with an infinite dimensional gauge symmetry (Lorentz groups of local coordinate transformations in every point of spacetime) \cite{tHooft1974,Deser1974}). It naturally makes sense to consider significantly simplified models of the same phenomenon which we described above by making dimensionality of the symmetry group finite and recall how Parisi-Sourlas dimensional reduction (due to the presence of random observer networks) emerges in this class of theories.

The first analyzed model of interest is a non-renormalizable scalar field theory with $Z_{2}$ global symmetry in $D=5$ and $6$ spacetime dimensions, with the Lagrangian density of the form ${\cal L}=\frac{1}{2}(\partial\phi)^{2}-\frac{1}{2}m^{2}\phi^{2}-\frac{1}{4}\lambda_{0}\phi^{4}-\ldots$, where $\ldots$ denotes higher-order terms in powers of the scalar field $\phi$ and its spacetime derivatives $\partial$. It is well-known that this theory is trivial \cite{Aizenman1982,Aizenman1983}, which implies that all its critical exponents coincide with the ones given by the mean field theory approximation (with logarithmic corrections
in 4 dimensions \cite{Aizenman1983}), i.e., if the number of spacetime dimensions $D>4$, the quantum effective action of the $\phi^{4}$ theory in the continuum limit can be well described by the one of a free massive scalar field theory with effective mass of the field being a known function of the bare coupling $\lambda_{0}$.

Consider a system of von Neumann detectors excited during the interaction events with quanta of the field $\phi$ \cite{Zurek1981,Zurek2003}. As usual, such detectors with monopole moments $J_{i}=J_{i}(t,x)$ can be modeled by terms in the Lagrangian density of the theory linear in the field variable $\phi$ as
\[
Z=\int{\cal D}\phi\exp\left(-i\sum_j\int d^{D}x\,\Big(\,\frac{1}{2}(\partial\phi)^{2}-
\frac{1}{2}m_{0}^{2}\phi^{2}-\frac{1}{4}\lambda_{0}\phi^{4}+J_{j}\phi\,\Big)\right),
\]
where the sum $\sum_{i}$ runs over detectors distributed in the world volume of the theory. It is often convenient to think of sources $J_{i}$ as a second, extra massive scalar field (with a suppressed kinetic term). We are specifically interested in the case where a very large number of such von Neumann detectors randomly located in the world volume of the theory is present with random couplings $J_{i}$ to the field $\phi$.

It can be seen straightforwardly that the physical setup described here is equivalent to the one realized in a quantum $\lambda\phi^{4}$ theory in a random external field or its discrete version, the random field $D$-dimensional Ising model (RFIM) well studied in literature, see for example \cite{Fytas2013,Fytas2016,Fytas2017,Fytas2017a}). A celebrated result by Parisi and Sourlas \cite{Parisi1979} states that the infrared behavior of RFIM is equivalent to the one of a similar theory (Ising model) in the absence of random external field but living in $(D-2)$ dimensions: namely, the most infrared divergent terms present in the perturbative expansion of the generating functionals of the two theories ($(D-2)$-dimensional IM and D-dimensional RFIM) coincide term by term. While the Parisi-Sourlas correspondence for Ising model breaks down for $D<3$ \cite{Fytas2013}, it has been shown to hold universally for $D\ge4$.

Of especial interest for us is the observation that the presence of a large number of von Neumann detectors drastically changes the structure of the Hilbert space of the theory, which for $D=4$ and $5$ can no longer be approximated by the mean field-theoretic partition function once the disorder associated with the external field is averaged out (as reflected in the change of critical exponents of the theory as well as correlation functions of the field).

The magnitude of the change in the structure of Hilbert space of the theory can be assessed by numerical simulations, which have been recently done in \cite{Fytas2013} for $D=3$ random field Ising model (RFIM), in \cite{Fytas2016,Fytas2017} for $D=4$ RFIM and in \cite{Fytas2017a} for $D=5$ RFIM. We have also performed lattice numerical simulations of RFIM and reproduced the known results for comparison of $D=4,5$ RFIM with pure Ising model in $D=2,3$
dimensions. Similar to \cite{Fytas2013,Fytas2016,Fytas2017} we have exploited the fact that RFIM achieves phase transition at zero temperature \cite{Fisher1986} and as such, it is sufficient to focus on the physics of the ground state of the theory. For numerical simulations, we have used minimum cost-flow algorithm \cite{AnglesdAuriac1985,Goldberg1987}.

In addition, we have also performed simulations of $D=6$ RFIM and compared its behavior with the one of pure Ising model in $D=4$ dimensions. As was expected, Parisi-Sourlas dimensional reduction was observed in $5$-dimensional and $6$-dimensional RFIM, with the critical exponents of $D=4$ RFIM deviating from the ones of $2$-dimensional Ising model due to the known breakdown of dimensional reduction mechanism in lower dimensions. Estimating critical exponents for $D=6$ RFIM we were unable to detect logarithmically weak corrections to the mean field approximation.

To confirm universality of Hilbert space restructuring in quantum field theories due to the presence of networks of observers/observation events, we have also performed numerical simulations of $Z_{2}$ gauge theory in $D=2+1$ and $3+1$ spacetime dimensions \cite{Balian1975,Creutz1980,Kehl1988} as well as $Z_{2}$ gauge theory in the presence of the random network of observers measuring gauge invariant quantities in $D=4+1$ and $5+1$ spacetime dimensions (here $+1$ denotes the dimension with periodic boundary conditions). Von Neumann observers were modeled by a scalar degree of freedom coupled to the $Z_{2}$ gauge field with the resulting free energy of the theory given by
\begin{equation}
F=\sum_{i,j,k,l}\sigma_{ij}\sigma_{jk}\sigma_{kl}\sigma_{li}+\sum_{i,j,n}g_{n}\tau_{i}\sigma_{ij}\tau_{j},
\label{eq:Z2observerTerms}
\end{equation}
where the couplings $g_{n}$ (strengths of detectors' couplings to $Z_{2}$ gauge field) and locations of insertions of the quenched disorder elements (observation events) were considered random and Gaussian-distributed. Again, we have observed effective dimensional reduction in the random field $Z_{2}$ gauge theory implying universality of this phenomenon across a wide range of theories with global and gauge symmetries.

\section{Numerical simulations of field theories\label(Sec:NumericsFieldTheories)}
\subsection{Lattice simulations of Ising model and random field Ising model (RFIM) }
The Ising model approximates (Euclidean) $\phi^{4}$ quantum field theory in the continuum limit (achieved for RFIM at zero temperature \cite{Fisher1986}).  Lattice simulations of zero-temperature RFIM in $D=4,5$ and $6$ dimensions were performed on hypercubic lattices with sizes $L=8,10,12,16$ and $20$. Ground states of the resulting IMs were calculated for $10^{6}$ realizations of disorder. For both IM and RFIM, finite-size scaling effects were taken into account. After extraction of $L$-dependence, the values of critical exponents were determined by extrapolating $L^{-1}\to0$. We obtained $\eta=0.1942\pm0.0022$, $\nu=0.8726\pm0.0182$ for $D=4$, $\eta=0.0442\pm0.0032$, $\nu=0.6293\pm0.0030$ for $D=5$ and $\eta=0.0103\pm0.0041$, $\nu=0.4892\pm0.0171$ for $D=6$.
\subsection{Lattice simulations of pure and random field $Z_2$ gauge field theories}
Monte-Carlo lattice simulations of Euclidean $Z_{2}$ and RF (random field) $Z_{2}$ gauge field theories were performed on periodic hypercubic lattices of the size $L=8,12,18,24$ and $28$ for the spatial part and fixed $L=2$ for the inverse temperature part of the lattice. For the RF $Z_{2}$ gauge field theory, $10^{6}$ realizations of random disorder were used. For $Z_{2}$ gauge field theory, we obtained $\beta=0.13\pm0.02$, $\nu=0.99\pm0.03$ for $D=2+1$ and $\beta=0.33\pm0.01$, $\nu=0.63\pm0.03$ for $D=3+1$ dimensions. (As usual, $+1$ denotes a dimension with periodic Matsubara boundary conditions.) For RF $Z_{2}$ gauge field theory, we found $\beta=0.11\pm0.03$, $\nu=0.65\pm0.04$ for $D=4+1$ and $\beta=0.30\pm0.05$, $\nu=0.65\pm0.04$ for $D=5+1$ dimensions.

\section{Deriving Starobinsky-Fokker-Planck equation using path integral approach\label{sec:SFP}}
In this Appendix, to illustrate the power of path integral approach for analyzing infrared dynamics of the scalar field 
in quasi-de Sitter universe, we shall derive the standard inflationary Fokker-Planck equation in the one-noise model. 
As usual we start with
\begin{equation}
\frac{\partial\Phi}{\partial{}N}=-\frac{1}{3H^2}\frac{\partial{}V}{\partial\Phi}+\frac{f}{H},
\end{equation}
where the noise $f=H\sigma$ possesses the correlation properties
\begin{equation}
\langle\sigma{}(N)\sigma(N')\rangle=\frac{H^2}{4\pi^2}\delta{}(N-N')
\end{equation}
(this equation is derived straightforwardly using the approach described in \cite{Starobinsky1988} under the assumption of vanishing slow roll parameters
$\epsilon_H,\,\eta_H\to{}0$). 
The partition function of the effective IR theory thus has the form
\begin{equation}
Z=\int{\cal D}\Phi{\cal D}\sigma\delta{}\left(\frac{\partial\Phi}{\partial{}N}+\frac{1}{3H^2}\frac{\partial{}V}{\partial\Phi}-\sigma\right)
\exp{}\left(-\int dN \frac{2\pi^2}{H^2}\sigma^2\right).\nonumber
\end{equation}
Introducing a Lagrangian multiplier for the functional delta function, integrating out the noise $\sigma$ as well as the Langrangian multiplier, we 
obtain
\begin{equation}
Z=\int{\cal D}\Phi\exp\left(-\int{}dN\frac{2\pi^2}{H^2}\left(\frac{\partial\Phi}{\partial{}N}+\frac{1}{3H^2}\frac{\partial{}V}{\partial\Phi}\right)^2\right).
\label{eq:1NoiseZ}
\end{equation} 
Thus the Lagrangian of the theory is
\begin{equation}
{\cal L}=\frac{2\pi^2}{H^2}\left(\frac{\partial\Phi}{\partial{}N}+\frac{1}{3H^2}\frac{\partial{}V}{\partial\Phi}\right)^2.
\label{eq:1NoiseL}
\end{equation}
The momentum conjugate to $\Phi$ is given by
\begin{equation}
P_\Phi=\frac{\partial{\cal L}}{\partial\Phi'}=\frac{\pi^2}{H^2}\left(\frac{\partial\Phi}{\partial{}N}+\frac{1}{3H^2}\frac{\partial{}V}{\partial\Phi}\right),
\label{eq:1NoiseMomentum}
\end{equation}
and the Hamiltonian of the theory corresponding to the Lagrangian (\ref{eq:1NoiseL}) is
\begin{equation}
H_\Phi=P_\Phi\Phi'-{\cal L}=P_\Phi\Phi'-{\cal L}=\frac{1}{8\pi^2}P_\Phi{}H^2{}P_\Phi-\frac{P_\Phi}{3H^2}\frac{\partial{}V}{\partial\Phi}.
\label{eq:1NoiseHamiltonian}
\end{equation}
The Starobinsky-Fokker-Planck equation \cite{Starobinsky1988} describing IR inflationary dynamics is obtained using this Hamiltonian and replacing $P_\Phi\to-\frac{\partial}{\partial\Phi}$ in the same fashion as Schroedinger equation is derived from the Feynman path integral for quantum mechanics:
\begin{equation}
\frac{\partial{}P(\Phi,N)}{\partial{N}}=H_\Phi\left(-\frac{\partial}{\partial{}\Phi},\Phi\right)P(\Phi,N).
\end{equation}
\section{Non-locality in the one-noise model\label{sec:NonLocal}}
\subsection{Useful preliminary expressions and used notations}
\subsubsection{Slow roll parameters}
In what follows, we consider the case of a single scalar field with a potential $V(\phi)$ propagating in a FRW spacetime with metric $ds^{2}=dt^{2}-a^{2}(t)d{\bf x}^{2}=a^{2}(t)(d\eta^{2}-d{\bf x}^{2})$. Ignoring dependence on spatial coordinates ${\bf x}$, slow roll parameters are defined according to the usual prescription
\begin{equation}
\epsilon_{H}=\frac{M_{P}^{2}}{4\pi}\left(\frac{dH/d\phi}{H}\right)^{2},
\label{eq:SlowRollEpsilonH}
\end{equation}
\begin{equation}
\eta_{H}=\frac{M_{P}^{2}}{4\pi}\frac{d^{2}H/d\phi^{2}}{H}.
\label{eq:SlowRollEtaH}
\end{equation}
Using the Hamilton-Jacobi equation for inflation
\begin{equation}
\left(\frac{dH}{d\phi}\right)^{2}-\frac{12}{M_{P}^{2}}H^{2}
=\frac{32\pi^{2}}{M_{P}^{4}}V(\phi)
\label{eq:HamiltonJacobi}
\end{equation}
and expressions
\begin{equation}
\frac{d\phi}{dt}=-\frac{M_{P}^{2}}{4\pi}\frac{dH}{d\phi},
\,\,\frac{dH}{dt}=-\frac{4\pi}{M_{P}^{2}}\left(\frac{d\phi}{dt}\right)^{2}
\label{eq:TimeDerRelations}
\end{equation}
one can demonstrate that
\begin{equation}
\epsilon_{H}=-\frac{dH/dt}{H^{2}}.
\label{eq:SlowRollEpsilonHTime}
\end{equation}
Using slow roll parameters $\epsilon_{H}$, $\eta_{H}$ rather than the usual slow roll parameters $\epsilon_{V}$, $\eta_{V}$ is more convenient as the end of inflationary stage corresponds to the condition $\epsilon_{H}=1$ being held exactly.

Other useful formulae which we use below include
\[
\frac{\partial(aH)}{\partial\eta}=(aH)^{2}(1-\epsilon_{H}),
\]
\begin{equation}
\frac{\partial\epsilon_{H}}{\partial\eta}=-2(\epsilon_{H}-\eta_{H})\epsilon_{H}aH,
\label{eq:DerivativeSlowRollPar}
\end{equation}
\[
\frac{\partial^{2}(aH)}{\partial\eta^{2}}=
(1-2\epsilon_{H}+2\epsilon_H^2-\epsilon_{H}\eta_{H})(aH)^{3}.
\]
The formula (\ref{eq:DerivativeSlowRollPar}) (rewritten in terms of inflationary efoldings $dN=Had\eta$) shows that taking time derivatives of slow roll parameters produces terms of higher order in slow roll expansion.
\subsubsection{Number of inflationary efoldings}
The Langevin and Fokker-Planck equations derived below are written
in terms of the number of efoldings $N=\log a$ rather than the world
time $t$ or conformal time $\eta$; it is therefore appropriate to
introduce the Jacobians associated with the corresponding change of
variables. We find:
\[
\frac{\partial}{\partial\eta}=aH\frac{\partial}{\partial N}.
\]
A number of useful formulae which will be used in later derivations
follow
\[
\epsilon_{H}=-\frac{1}{H}\frac{dH}{dN},\,\,\eta_{H}=\epsilon_{H}-
\frac{M_{P}}{\sqrt{16\pi}}\frac{d\epsilon_{H}/d\phi}{\sqrt{\epsilon_{H}}},
\]
\[
\frac{a''}{a}=(2-\epsilon_{H})(Ha)^{2},
\]
\[
\frac{\partial}{\partial N}(aH)=aH(1-\epsilon_{H}).
\]
\subsection{Separating scalar field into IR and UV parts. Langevin equation}
The separation of the field into subhorizon and superhorizon parts
is done according to
\begin{equation}
\phi=\Phi+\frac{1}{(2\pi)^{3/2}}\int d^{3}k\,\theta(k-\delta aH)\left(a_{k}\phi_{k}(\eta)
e^{-i{\bf kx}}+{\rm h.c.}\right),
\label{eq:IRUVSeparation}
\end{equation}
where as usual $\theta(\ldots)$ is the Heaviside step function of the argument and $\delta$ is a free dimensionless parameter identified with IR/UV separation scale (usually, in stochastic formalism it is taken to be small, $\delta\ll1$, but not too small in order for potential terms to remain sub-dominant). Substituting the expression into the operator equation $\Box\phi+\frac{\partial V}{\partial\phi}=0$ and neglecting potential term for the UV part of the field, we obtain:
\begin{equation}
\Box\Phi+\frac{1}{a^{3}}(u''-\nabla^{2}u-\frac{a''}{a}u+\ldots)
\approx-\frac{\partial V}{\partial\Phi},
\label{eq:BoxEqSeparation}
\end{equation}
where $\ldots$ denotes terms related to the potential and
\[
\Box\Phi=H^{2}\frac{\partial^{2}\Phi}{\partial N^{2}}+H^{2}(3-\epsilon_{H})
\frac{\partial\Phi}{\partial N}-\frac{\nabla^{2}\Phi}{a^{2}}
\]
and
\[
u=\frac{a}{(2\pi)^{3/2}}\int d^{3}k\theta(k-\delta aH)\left(a_{k}u_{k}
(\eta)e^{-i{\bf kx}}+{\rm h.c.}\right).
\]
Substituting $u$ into the Eq. (\ref{eq:BoxEqSeparation}) and using the expression
\[
u''=-\frac{\delta^{2}}{(2\pi)^{3/2}}\int d^{3}k\delta'(k-\delta aH)(aH)^{4}
(1-\epsilon_{H})^{2}(a_{k}u_{k}e^{-i{\bf kx}}+{\rm h.c.})-
\]
\[
-\frac{2\delta}{(2\pi)^{3/2}}\int d^{3}k\delta(k-\delta aH)(aH)^{3}
(1-2\epsilon_{H}+2\epsilon_H^2-\epsilon_{H}\eta_{H})(a_{k}u_{k}e^{-i{\bf kx}}+{\rm h.c.})-
\]
\[
-\frac{2\delta}{(2\pi)^{3/2}}\int d^{3}k\delta(k-\delta aH)(aH)^{2}
(1-\epsilon_{H})(a_{k}u_{k}'e^{-i{\bf kx}}+{\rm h.c.})+
\]
\[
+\frac{1}{(2\pi)^{3/2}}\int d^{3}k\theta(k-\delta aH)(a_{k}u_{k}''e^{-i{\bf kx}}+{\rm h.c.}),
\]
we finally obtain the ``Langevin'' equation for the infrared part of the field
\begin{equation}
\frac{\partial^{2}\Phi}{\partial N^{2}}+(3-\epsilon_{H})\frac{\partial\Phi}{N}
-\frac{\nabla^{2}\Phi}{(aH)^{2}}+\frac{\partial^{2}V}{\partial\Phi^{2}}=
f_{1}(t,{\bf x})+f_{2}(t,{\bf x}),
\label{eq:LangevinEq}
\end{equation}
where the ``noise'' operators $f_{1}$ and $f_{2}$ are defined according to
\begin{equation}
f_{1}=-\frac{\delta^{2}aH^{2}(1-\epsilon_{H})^{2}}{(2\pi)^{3/2}}\int
d^{3}k\delta'(k-\delta aH)(a_{k}u_{k}e^{-i{\bf kx}}+{\rm h.c.}),
\label{eq:Noisef1}
\end{equation}
\[
f_{2}=-\frac{2\delta}{(2\pi)^{3/2}}\int d^{3}k\delta(k-\delta aH)
\left[H(1-2\epsilon_{H}+2\epsilon_H^2-\epsilon_{H}\eta_{H})(a_{k}u_{k}e^{-i{\bf kx}}+{\rm h.c.})+\right.
\]
\begin{equation}
\left.\frac{1-\epsilon_{H}}{a}(a_{k}u_{k}'e^{-i{\bf kx}}+{\rm h.c.})\right].
\label{eq:Noisef2}
\end{equation}
The part of the noise term $f_{1}$ seems to be suppressed at small $\delta\ll1$. However, as we shall see, generally this is not the case as powers of $\delta$ are canceled out in the observable quantities.
\subsection{Commutation relations of the noise operators $f_1$ and $f_2$}
While the noise terms commute with terms on the r.h.s. of the Eq. (\ref{eq:LangevinEq}), it is also useful to check their self-commutation relations (due to ultra-locality in the quasi-de Sitter regime, we shall be particularly interested in commutation relations of operators at the same spatial point ${\bf x}$). We find that
\begin{equation}
[f_{1}(N),f_{1}(N')]=0,
\label{eq:Commutationf1f1}
\end{equation}
\begin{equation}
[f_{2}(N),f_{2}(N')]=0,
\label{eq:Commutationf2f2}
\end{equation}
\begin{equation}
[f_{1}(N),f_{2}(N')]=-\frac{i\delta^{3}H^{2}(1-\epsilon_{H})}{\pi^{2}}\delta'(N-N').
\label{eq:Commutationf1f2}
\end{equation}
It is worth noting that although the Langevin equation (\ref{eq:LangevinEq}) is considered to be quasi-classical, operators $f_{1}$and $f_{2}$ are not generally commuting although their commutator is small at $\delta\ll1$ and becomes vanishing by the end of inflation when $\epsilon_{H}\to1$.

This is in contrast with the Langevin-Starobinsky equation \cite{Starobinsky1988} for a scalar field on a fixed de Sitter background
\begin{equation}
\dot{\Phi}+\frac{1}{3H_{0}}\frac{\partial V}{\partial\Phi}=f,
\label{eq:StarobinskyLangevin}
\end{equation}
where
\begin{equation}
f=-\frac{\delta aH_{0}^{3}i}{4\pi{}^{3/2}}\int d^{3}k\delta(k-\delta aH_{0})
\frac{1}{k^{3/2}}\left(a_{k}e^{-i{\bf kx}}-a_{k}^{\dagger}e^{i{\bf kx}}\right).
\label{eq:StarobinskyNoise}
\end{equation}
 One can immediately see that due to antisymmetric form of the combination $\left(a_{k}e^{-i{\bf kx}}-a_{k}^{\dagger}e^{i{\bf kx}}\right)$ the
noise term $f(t)$ commutes with itself if the same spatial point is considered:
\[
[f(t,{\bf x}),f(t',{\bf x})]=0.
\]
For points with large spatial (superhorizon) separation the noise terms do not commute even in this simplified case:
\[
[f(t,{\bf x}),f(t',{\bf x}')]=\frac{H^{3}}{2\pi^{2}}\sin(\epsilon aH|{\bf x}-{\bf x}'|)\delta(t-t').
\]
The reason of this discrepancy with our result is due to dropping terms $\sim f_{1}$ as suppressed by additional powers of $\delta$ at $\delta\ll1$ during the derivation of (\ref{eq:StarobinskyNoise}); on the other hand, when deriving (\ref{eq:Commutationf1f1}) - (\ref{eq:Commutationf1f2}) all terms are kept explicitly. It is thus useful to remember that the quantum nature of the noise in the Langevin equation (\ref{eq:LangevinEq}) is not eradicated completely during the quasi-de Sitter inflationary stage when $\epsilon_{H}\ll1$.
\subsection{Correlation functions of the noise operators $f_{1}$ and $f_{2}$}
While we consider the field operators $f_{1}(N,{\bf x})$ and $f_{2}(N,{\bf x})$ quasi-classical quantities (based on their commutation relations in the regime $\delta\ll1$ as well as their commuting with other terms in the Langevin equation (\ref{eq:LangevinEq})) in what follows, it is necessary to determine their stochastic properties. Those are given by the expectation values in the vacuum state of the Fock space of modes $u_{k}$. We find:
\[
\langle f_{2}(N,{\bf x})f_{2}(N',{\bf x})\rangle=\frac{2\delta^{2}H}{\pi^{2}}
\left[\left(\frac{1-2\epsilon_{H}+2\epsilon_H^2-\epsilon_{H}\eta_{H}}{\sqrt{1-\epsilon_{H}}}{\rm Re}u_{k}
+\frac{\sqrt{1-\epsilon_{H}}}{Ha}{\rm Re}u_{k}'\right)^{2}+\right.
\]
\begin{equation}
\left.+\left(\frac{1-2\epsilon_{H}+2\epsilon_H^2-\epsilon_{H}\eta_{H}}{\sqrt{1-\epsilon_{H}}}{\rm Im}u_{k}
+\frac{\sqrt{1-\epsilon_{H}}}{Ha}{\rm Im}u_{k}'\right)_{k=\delta aH}^{2}\right]\delta(N-N'),
\label{eq:f2f2corr}
\end{equation}
\[
\langle f_{1}(N,{\bf x})f_{1}(N',{\bf x})\rangle=-\frac{H^{2}\delta^{2}}{2\pi^{2}}
\left[(1-\epsilon_{H})\delta aH(u_{k}u_{k}^{*})_{k=\delta aH}\delta''(N-N')+\right.
\]
\[
+2(1-\epsilon_{H})^{2}\delta aH(u_{k}u_{k}^{*})_{k=\delta aH}\delta'(N-N')+
\]
\begin{equation}
\left.+(1-\epsilon_{H})^{2}(\delta aH)^{2}\left(
\frac{d}{dk}(u_{k}u_{k}^{*})\right)_{k=\delta aH}\delta'(N-N')\right],
\label{eq:f1f1corr}
\end{equation}
\[
\langle f_{1}(N,{\bf x})f_{2}(N',{\bf x})\rangle=\frac{\delta^{3}H^{2}}{\pi^{2}}
\left[aH(1-2\epsilon_{H}+2\epsilon_H^2-
\epsilon_{H}\eta_{H})(u_{k}u_{k}^{*})_{k=\delta aH}+\right.
\]
\begin{equation}
\left.+(1-\epsilon_{H})(u_{k}u_{k}^{*}{}')_{k=\delta aH}\right]\delta'(N-N').
\label{eq:f1f2corr}
\end{equation}
We emphasize that the expressions (\ref{eq:f2f2corr})-(\ref{eq:f1f2corr}) are exact to all orders in slow roll parameters $\epsilon_{H}$, $\eta_{H}$.
They can be significantly simplified if the leading order in slow roll parameters is kept; in the regime $\eta_{H}\to0$, $\epsilon_{H}\to0$ we have
\[
u_{k}\approx\frac{1}{\sqrt{2k}}\left(\frac{i}{k\eta}-1\right)e^{-ik\eta}=
-\frac{1}{\sqrt{2k}}\left(\frac{iHa}{k}+1\right)e^{\frac{ik}{Ha}},
\]
\[
u_{k}'\approx\frac{1}{\sqrt{2k}\eta}e^{-ik\eta}\left(-\frac{i}{k\eta}-ik\eta+1\right)=
-\frac{Ha}{\sqrt{2k}}e^{\frac{ik}{Ha}}\left(\frac{iHa}{k}+\frac{ik}{Ha}+1\right),
\]
and
\begin{equation}
\langle f_{2}(N,{\bf x})f_{2}(N',{\bf x})\rangle=\frac{4H^{2}}{\pi^{2}}\delta(N-N'),
\label{eq:f2f2Star}
\end{equation}
\begin{equation}
\langle f_{1}(N,{\bf x})f_{1}(N',{\bf x})\rangle=
\frac{H^{2}}{8\pi^{2}}\delta'(N-N')-\frac{H^{2}}{4\pi^{2}}\delta''(N-N'),
\label{eq:f1f1Star}
\end{equation}
\begin{equation}
\langle f_{1}(N,{\bf x})f_{2}(N',{\bf x})\rangle=\frac{H^{2}}{\pi^{2}}\delta'(N-N').
\label{eq:f1f2Star}
\end{equation}
We are thus forced to conclude that in the general case the one noise model produces \emph{a non-local} 
effective theory (and a rather hard one to deal with), which can immediately be seen from the behavior of 
the correlation functions (\ref{eq:f2f2Star})-(\ref{eq:f1f2Star}) as well as after integrating the noise $f_1\, f_2$ out in the partition function of the theory. This non-locality cannot be really neglected as the correlation functions of the noise $~\delta'(N-N'),\, \delta''(N-N')$ are not suppressed. It also
hints on the presence of an additional stochastic degree of freedom which was integrated out to produce the non-local behavior and forces us to apply the two-noise model described in the main text.
\subsubsection{Relation to the Starobinsky's stochastic formalism}
Two observations are in order. First of all, we note that even keeping $f_2$ only (which is essentially equivalent to
using the approximation $\epsilon_H\ll1$, $\delta\ll1$ employed in \cite{Starobinsky1988} and throughout the literature)
we do not reproduce Starobinsky's result for the numerical factor in front of the correlation function (\ref{eq:f2f2Star}) --- the difference between the two results is a factor of $3/4$ (which is crucial given that it determines the correct value for de Sitter entropy!).
To understand what happens, let us recall how it is derived. If all terms suppressed by higher powers of slow roll parameters
are neglected, the resulting equation for the superhorizon part of the field has the form
\[
\frac{\partial \Phi}{\partial t}=-\frac{1}{3H}\frac{\partial{}V}{\partial\Phi}+f,
\]
where
\[
f=-\frac{i\delta aH^3}{4\pi^{3/2}}\int d^3k\delta(k-\delta aH)\frac{1}{k^{3/2}}
\left(a_ke^{-i\bf{kx}}-a_k^\dagger{}e^{-i\bf{kx}}\right). 	
\]
Note however that the terms $\ddot{\phi}$ which we neglected in the equations above would also contain the contribution $\sim{}\dot{f}$.
While most contributions to $\dot{f}$ are suppressed by additional powers of $\delta$, there is also a contribution present
which is proportional to $\sim{}Hf$. This contribution is exactly the one which accounts for the difference between our and Starobinsky's result.
However, keeping terms like this, we should be extra careful since $f$ is a stochastic variable, which we are trying to differentiate.

Second, we note that the correlators \ref{eq:f1f2corr} and \ref{eq:f1f1corr} are \emph{not} suppressed by powers of $\delta$ and thus should generally be kept. The resulting theory (after the noise term $f1$ is integrated out) is non-local in $N$. This non-locality hints on an existence of an additional effective field variable which has been integrated out to obtain the resulting non-local theory.

 \acknowledgments{The work of A.O.B. was supported by the RFBR grant No.20-02-00297 and by the Foundation for 
 Theoretical Physics Development ``Basis''.}


\end{document}